# Coupled-mode theory for non-periodic structured waveguides

M.I. Ayzatsky[1]

National Science Center Kharkov Institute of Physics and Technology (NSC KIPT), 61108, Kharkov, Ukraine

In this work a new generalization of the theory of coupled modes for non-periodic structured waveguide is presented. Based on a set of eigen waves of a homogeneous periodic waveguide, a new basis of vector functions is introduced that takes into account the non-periodicity of the waveguide. Representing the total field as the sum of these functions with unknown scalar coefficients, a system of coupled equations that determines the dependence of these coefficients on the longitudinal coordinate has been obtained. It was shown that the single-wave equation has an additional "phase" term. In the frame of proposed approach, the z-dependent series impedance and local wave vector were introduced for structured waveguides

## 1. INTRODUCTION

Waveguides that consist of similar (but not always identical) cells are called structured. Structured waveguides based on coupled resonators play an important role in many applications. Their use in active devices, in comparison with passive ones, has a number of distinctive features. First, TH modes with a large longitudinal component of the electric field are used. Secondly, in accelerators and high-frequency electronic devices, the spatial distribution of longitudinal electromagnetic fields (especially their phases) over the structure plays a decisive role. Thirdly, the active elements of these devices are electron beams.

There are powerful programs that can be used to calculate the RF characteristics of structured waveguides and the electromagnetic fields induced in them by electron beams, but they are difficult to use for characterization and preliminary design. For these purposes simpler approaches are needed.

Electromagnetic fields in the periodic structured waveguides can be effectively described by the Floquet-Bloch's theory and the electrodynamic approach based on the field expansion in forward and backward waves which constitute a complete orthogonal set of vector functions and have a rigorous physical basis. In this case the problem is one-dimensional, since the wave amplitudes depend only on one coordinate and can be found by solving a set of ordinary differential equations that are not coupled.

In the general case, when the coupled resonators are different, there are no physical concepts that could simplify the understanding of the electromagnetic process. If the parameters of resonators change smoothly and slowly, we can expect that the electromagnetic fields will have some features of forward and backward waves. In the case of inhomogeneous smooth waveguides, approximate approaches are a powerful tool for study its properties [1,2]. The development of approximate approaches for structured waveguides is at an early stage. A method based on a generalization of the theory of coupled modes for the case when the structure and the non-periodicity can be described by differential operators was proposed [3,4]. To use this approach for a waveguide whose structure is determined by the boundaries, it is necessary to transform the side walls into a smooth cylinder and obtain differential equations describing the fields in a smooth waveguide with inhomogeneous filling. This is a complex (and in some cases impossible) procedure.

In this work we presented a new generalization of the theory of coupled modes for non-periodic structured waveguide with ideal metal walls.

## 2. BASIC EQUATIONS

Let us first consider a periodic structured waveguide with metal walls. We will consider the axisymmetric waveguides.

In most cases the boundary of a periodic structured waveguide is completely determined by a finite set of geometrical parameters $g_i$, $i=1,...,I$. For example, for a circular disk-loaded waveguide (see Figure 1), we have four geometrical parameters: $g_1 = b$ – the radius of the waveguide, $g_2 = a$ - the radius of the aperture, $g_3 = t$ - the thickness of the disk and $g_4 = d$ – the distance between disks (waveguide period $D = t + d$). We can write the dependence of the radius of this waveguide on the longitudinal coordinate $z$ as

$$R(z) = \begin{cases} g_2, & (n-1)D < z < (n-1)D+t, \\ g_1, & (n-1)D+t < z < nD. \end{cases} \quad (1),$$

For each fixed $z$ there are a subset of $g_i^{(l)}$ (let's call them local geometrical parameters), that determine the geometry of the cross section $S_\perp(z, g_i^{(l)})$ which can have a complex shape, even be multiply connected. The remaining parameters will be called global $g_i^{(g)}$. The division of the geometrical parameters $g_i$ into local and global ones depends on $z$. For example, in the case of a circular disk-loaded waveguide (see Figure 1) we have $g_2^{(l)} = a$ for $(n-1)D < z < (n-1)D+t$ and $g_1^{(l)} = b$ for $(n-1)D+t < z < nD$. There may be restrictions on the range of variation of each parameter $g_i$, determined by the geometry of the waveguide.

---

[1] E-mail: aizatsky@kipt.kharkov.ua, mykola.aizatsky@gmail.com

We will assume that all quantities have a time variation given by $\exp(-i\omega t)$. The behavior of electromagnetic field is governed by Maxwell's equations

$$rot\,\vec{E} = i\omega\mu_0\vec{H}, \tag{2}$$

$$rot\,\vec{H} = -i\omega\varepsilon_0\varepsilon\vec{E} + \vec{j}. \tag{3}$$

For periodic waveguide without losses we have two sets of eigen waves $\{\vec{E}_s(\vec{r}),\vec{H}_s(\vec{r})\} = \{\vec{\tilde{E}}_s(\vec{r}),\vec{\tilde{H}}_s(\vec{r})\}\exp(\gamma_s z)$ and $\{\vec{E}_{-s}(\vec{r}),\vec{H}_{-s}(\vec{r})\} = \{\vec{\tilde{E}}_{-s}(\vec{r}),\vec{\tilde{H}}_{-s}(\vec{r})\}\exp(\gamma_{-s} z)$, $\gamma_{-s} = -\gamma_s$, which are the solutions to such equations ($s > 0$ - forward waves, $s < 0$ - backward waves)

$$\begin{aligned} rot\,\vec{E}_s &= i\omega\mu_0\vec{H}_s, \\ rot\,\vec{H}_s &= -i\omega\varepsilon_0\varepsilon\vec{E}_s, \end{aligned} \tag{4}$$

satisfy the orthogonality condition

$$N_{s,s'} = \int_{S_\perp(z)} \left\{\left[\vec{E}_s\vec{H}_{s'}\right] - \left[\vec{E}_{s'}\vec{H}_s\right]\right\}\vec{e}_z dS = \begin{cases} 0, & s' \neq -s, \\ N_s, & s' = -s, \end{cases} \tag{5}$$

and the boundary conditions on the side metallic surface of the waveguide

$$\begin{aligned} \vec{E}_{s,\tau} &= 0, \\ \vec{H}_{s,\perp} &= 0. \end{aligned} \tag{6}$$

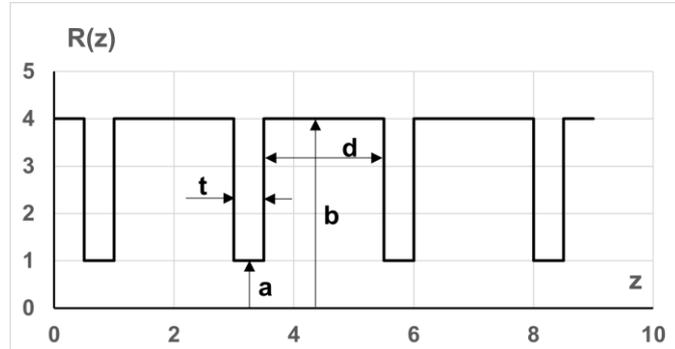

Figure 1

Eigenvectors $\{\vec{E}_s(\vec{r}),\vec{H}_s(\vec{r})\}$ depend on the values of geometric parameters $g_i$ and it can be assumed that this dependence can be considered as functional and written $\vec{E}_s = \vec{E}_s(\vec{r}_\perp, z, g_1,..., g_I)$, $\vec{H}_s = \vec{H}_s(\vec{r}_\perp, z, g_1,..., g_I)$. The domain of these functions is determined by the geometry of the waveguide. For example, the longitudinal component of the electric field in a cylinder resonator $E_z = J_0\left(\frac{\lambda_s}{b}r\right)\cos\left(\frac{\pi}{d}z\right)$ can be considered as a function of four variables $E_z = E_z(r, z, b, d)$, where $0 \leq r \leq b$, $0 \leq z \leq d$.

We can introduce new vector functions $\vec{E}_s^{(z)} = \vec{E}_s(\vec{r}_\perp, z, g_1^{(z)}(z),..., g_I^{(z)}(z))$, $\vec{H}_s^{(z)} = \vec{H}_s(\vec{r}_\perp, z, g_1^{(z)}(z),..., g_I^{(z)}(z))$, where $g_i^{(z)}(z)$ and its derivatives are continuous functions of $z$. The set $g_i^{(z)}(z)$ doesn't describe any real waveguide. But for each fixed $z$ the vectors $\vec{E}_s^{(z)}$, $\vec{H}_s^{(z)}$ represent the fields of a periodic waveguide (analogue of a virtual waveguide [3,4]) in the cross section $S_\perp^{(z)}(z, g_i^{(l,z)}(z))$, where $g_i^{(l,z)}(z)$ are the subset of local geometrical parameters.

The vector functions $\vec{E}_s^{(z)}$, $\vec{H}_s^{(z)}$ are no longer the solutions to equations (4). Indeed, as

$$\frac{\partial \vec{E}_s^{(z)}}{\partial z} = \frac{\partial \vec{E}_s^{(z)}}{\partial z}\Big|_{g_i=const} + \sum_i \frac{\partial \vec{E}_s^{(z)}}{\partial g_i^{(z)}}\frac{dg_i^{(z)}}{dz}, \tag{7}$$

then

$$\begin{aligned} rot\vec{E}_s^{(z)} &= rot\vec{E}_s^{(z)}\Big|_{g_i=const} + \vec{E}_s^{(\nabla)} = i\omega\mu_0\vec{H}_s^{(z)} + \vec{E}_s^{(\nabla)}, \\ rot\,\vec{H}_s^{(z)} &= rot\,\vec{H}_s^{(z)}\Big|_{g_i=const} + \vec{H}_s^{(\nabla)} = -i\omega\varepsilon_0\varepsilon\vec{E}_s^{(z)} + \vec{H}_s^{(\nabla)}, \end{aligned} \tag{8}$$

where

$$\vec{E}_s^{(\nabla)} = \sum_i \frac{dg_i^{(z)}}{dz} \left[ \vec{e}_z \frac{\partial \vec{E}_s^{(z)}}{\partial g_i^{(z)}} \right],$$

$$\vec{H}_s^{(\nabla)} = \sum_i \frac{dg_i^{(z)}}{dz} \left[ \vec{e}_z \frac{\partial \vec{H}_s^{(z)}}{\partial g_i^{(z)}} \right]. \qquad (9)$$

The new vector functions $\vec{E}_s^{(z)}$, $\vec{H}_s^{(z)}$ still obey the boundary conditions (6) on the contour $\vec{r}_\perp = \vec{r}_\perp^{(z)}(z, g_i^{(l,z)}(z))$, which limits the cross section $S_\perp^{(z)}(z, g_i^{(l,z)}(z))$ of some waveguide, and the orthogonality conditions

$$N_{s,s'}^{(z)} = \int_{S_\perp^{(z)}(z)} \left\{ \left[ \vec{E}_s^{(z)} \vec{H}_{s'}^{(z)} \right] - \left[ \vec{E}_{s'}^{(z)} \vec{H}_s^{(z)} \right] \right\} \vec{e}_z dS = \begin{cases} 0, & s' \neq -s, \\ N_s^{(z)}\left(g_i^{(z)}(z)\right), & s' = -s. \end{cases} \qquad (10)$$

As $\{\vec{E}_s(\vec{r}), \vec{H}_s(\vec{r})\} = \{\vec{\tilde{E}}_s(\vec{r}), \vec{\tilde{H}}_s(\vec{r})\} \exp(\gamma_s z)$ then $\{\vec{E}_z^{(z)}(\vec{r}), \vec{H}_z^{(z)}(\vec{r})\} = \{\vec{\tilde{E}}_z^{(z)}(\vec{r}), \vec{\tilde{H}}_z^{(z)}(\vec{r})\} \exp(\gamma_z^{(z)} z)$, where $\gamma_s^{(z)}(z) = \gamma_s\left(g_1^{(z)}(z), \ldots, g_I^{(z)}(z)\right)$.

Before using the vector functions $\vec{E}_s^{(z)}$, $\vec{H}_s^{(z)}$ to describe the non-periodic structured waveguides it is necessary to define a procedure for choosing functions $g_i(z)$. For simplicity, we consider this procedure for the case of a circular disk-loaded waveguide.

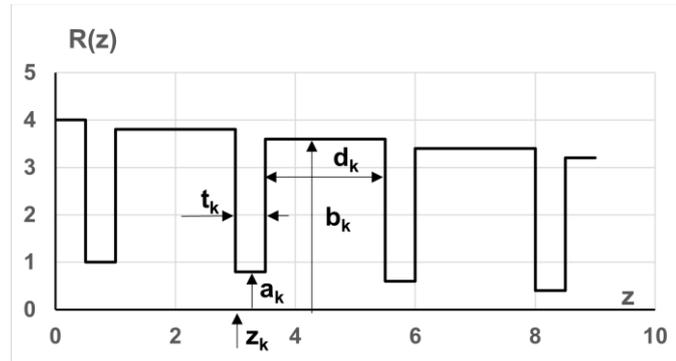

*Figure 2*

For non-periodic circular disk-loaded waveguide we will introduce the notion of the elementary cell with a number $k$ which starts at $z_k = \sum_i^{k-1}(t_i + d_i)$ and consists of a disk ($z_k < z < z_k + t_k$, thickness $g_{3,k} = t_k$, the radius of opening $g_{2,k} = a_k$) and the following segment of a circular waveguide ($z_k + t_k < z < z_k + t_k + d_k$, the radius $g_{1,k} = b_k$, length $g_{4,k} = d_k$) (see Figure 2). For this waveguide at $z_k < z \leq z_k + t_k$ there is one local parameter $g_{2,k}^{(l)} = a_k$, at $z_k + t_k < z \leq z_k + t_k + d_k$ the local parameter is $g_{1,k}^{(l)} = b_k$. The vector functions $\vec{E}_s^{(z)}$, $\vec{H}_s^{(z)}$ will fulfill the orthogonality conditions and the boundary conditions (6) on the surface of the considered waveguide when the value of corresponding function $g_i^{(l,z)}(z)$ equals the value of the local parameter $g_{i,k}^{(l)}$. The values of global parameters can be arbitrary.

For circular disk-loaded waveguide at $z_k < z < z_k + t_k$ the local parameter $g_{2,k}^{(l)}$ is constant $g_{2,k}^{(l)} = a_k$, hence the function should be constant too: $g_2^{(z)}(z) = a_k$, $z_k < z < z_k + t_k$. The function $g_1^{(z)}(z)$ can be arbitrary, but there must be continues transition (including a derivative) from $g_1^{(z)}(z_k - 0) = b_{k-1}$ to $g_1^{(z)}(z_k + t_k + 0) = b_k$. The functions $g_3^{(z)}(z), g_4^{(z)}(z)$ can be arbitrary along the whole $k$ cell, but there must be continues transition between $k-1, k, k+1$ cells. One set of the possible distributions are depicted in Figure 2. For transition regions, we used the following continuations

$$g_1^{(z)}(z) = b_{k-1} + (b_k - b_{k-1}) \sin\left\{ \frac{\pi}{2} \left( \frac{z - z_k}{t_k} \right)^2 \right\}, \quad z_k < z < z_k + t_k, \qquad (11)$$

$$g_2^{(z)}(z) = a_k + (a_{k+1} - a_k) \sin\left\{ \frac{\pi}{2} \left( \frac{z - z_k - t_k}{d_k} \right)^2 \right\}, \quad z_k + t_k < z < z_k + t_k + d_k. \qquad (12)$$

These functions ensure the continuity of functions and its derivatives. The values of global parameters $g_3^{(z)}(z), g_4^{(z)}(z)$ for the case under consideration are constant along $z$

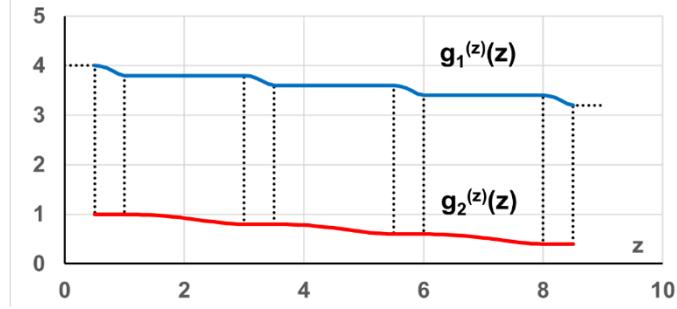

*Figure 3*

The sets of eigen solutions $\vec{E}_s, \vec{H}_s$ are complete. We can expect that the sets $\vec{E}_s^{(z)}$, $\vec{H}_s^{(z)}$ are complete, too, and we can look for a solution of equations (2) and (3) in the form of such a series

$$\vec{H}(\vec{r}) = \sum_{s>0} \left\{ C_s(z) \vec{H}_s^{(z)}(\vec{r}) + C_{-s}(z) \vec{H}_{-s}^{(z)}(\vec{r}) \right\}. \tag{13}$$

Using the vector relation

$$rot\left(C_s \vec{H}_s^{(z)}\right) = \left[\nabla C_s \vec{H}_s^{(z)}\right] + C_s \, rot \, \vec{H}_s^{(z)} \tag{14}$$

and assuming that the series (13) can be differentiated term by term, we get from (3)

$$\vec{E} = -\frac{1}{i\omega\varepsilon_0 \varepsilon} rot \, \vec{H} + \frac{\vec{j}}{i\omega\varepsilon_0 \varepsilon} =$$
$$\sum_s \left(C_s \vec{E}_s^{(z)} + C_{-s} \vec{E}_{-s}^{(z)}\right) - \frac{1}{i\omega\varepsilon_0 \varepsilon} \sum_s \left( \frac{dC_s}{dz}\left[\vec{e}_z \vec{H}_s^{(z)}\right] + \frac{dC_{-s}}{dz}\left[\vec{e}_z \vec{H}_{-s}^{(z)}\right] + C_s \vec{H}_s^{(\nabla)} + C_{-s} \vec{H}_{-s}^{(\nabla)} \right) + \frac{\vec{j}}{i\omega\varepsilon_0 \varepsilon}. \tag{15}$$

Suppose that

$$\sum_s \left( \frac{dC_s}{dz}\left[\vec{e}_z \vec{H}_s^{(z)}\right] + \frac{dC_{-s}}{dz}\left[\vec{e}_z \vec{H}_{-s}^{(z)}\right] \right) = \vec{j}_\perp - \sum_s \left(\vec{H}_s^{(\nabla)} C_s + \vec{H}_{-s}^{(\nabla)} C_{-s}\right), \tag{16}$$

where $\vec{j} = \vec{j}_z + \vec{j}_\perp$. We cannot use in (16) the full current $\vec{j}$ as the sums don't have the longitudinal components.

Then (15) takes the form

$$\vec{E} = \sum_s \left(C_s \vec{E}_s^{(z)} + C_{-s} \vec{E}_{-s}^{(z)}\right) + \frac{\vec{j}_z}{i\omega\varepsilon_0 \varepsilon}. \tag{17}$$

Substitution of (17) and (13) into the equation (2) gives

$$\sum_s \left( \frac{dC_s}{dz}\left[\vec{e}_z \vec{E}_s^{(z)}\right] + \frac{dC_{-s}}{dz}\left[\vec{e}_z \vec{E}_{-s}^{(z)}\right] \right) = -\frac{1}{i\omega\varepsilon_0 \varepsilon} rot \, \vec{j}_z - \sum_s \left(C_s \vec{E}_s^{(\nabla)} + C_{-s} \vec{E}_{-s}^{(\nabla)}\right). \tag{18}$$

Multiplying (18) by $\vec{H}_{-s'}^{(z)}$, (16) by $\vec{E}_{-s'}^{(z)}$, adding the resulting equations and integrating the result over the cross-section, we obtain

$$N_{s'}^{(z)} \frac{dC_{s'}}{dz} = \int_{S_\perp^{(z)}(z)} \left\{ \vec{j}_\perp \vec{E}_{-s'}^{(z)} - \frac{1}{i\omega\varepsilon_0 \varepsilon} \vec{H}_{-s'}^{(z)} rot \, \vec{j}_z - \vec{E}_{-s'}^{(z)} \sum_s \left(\vec{H}_s^{(\nabla)} C_s + \vec{H}_{-s}^{(\nabla)} C_{-s}\right) - \vec{H}_{-s'}^{(z)} \sum_s \left(\vec{E}_s^{(\nabla)} C_s + \vec{E}_{-s}^{(\nabla)} C_{-s}\right) \right\} dS, \tag{19}$$

where

$$N_s^{(z)} = \int_{S_\perp^{(z)}} \left\{ \left[\vec{E}_s^{(z)} \vec{H}_{-s}^{(z)}\right] - \left[\vec{E}_{-s}^{(z)} \vec{H}_s^{(z)}\right] \right\} \vec{e}_z dS. \tag{20}$$

Similarly, multiplying (18) by $\vec{H}_{s'}$, (16) by $\vec{E}_{s'}$, adding and integrating give

$$N_{s'}^{(z)} \frac{dC_{-s'}}{dz} = -\int_{S_\perp^{(z)}(z)} \left\{ \vec{j}^t \vec{E}_{s'}^{(z)} - \frac{1}{i\omega\varepsilon_0 \varepsilon} \vec{H}_{s'}^{(z)} rot \, \vec{j}^l - \vec{E}_{s'}^{(z)} \sum_s \left(\vec{H}_s^{(\nabla)} C_s + \vec{H}_{-s}^{(\nabla)} C_{-s}\right) - \vec{H}_{s'}^{(z)} \sum_s \left(\vec{E}_s^{(\nabla)} C_s + \vec{E}_{-s}^{(\nabla)} C_{-s}\right) \right\} dS. \tag{21}$$

We transform the integrand in (19).

$$\left(\vec{H}_{-s'}^{(z)} rot \, \vec{j}_z\right) = \vec{H}_{-s'}^{(z)} rot\left(j_z \vec{e}_z\right) = \vec{H}_{-s'}^{(z)} \left(\left[\nabla j_z \vec{e}_z\right] - j_z \, rot \, \vec{e}_z\right) =$$
$$= \vec{H}_{-s'}^{(z)} \left[\nabla j_z \vec{e}_z\right] = -\left[\nabla j_z \vec{H}_{-s'}^{(z)}\right] \vec{e}_z = -\vec{e}_z \left\{ \left[\nabla j_z \vec{H}_{-s'}^{(z)}\right] - j_z \, rot\left(\vec{H}_{-s'}^{(z)} |_{g_i=const}\right) + j_z \, rot\left(\vec{H}_{-s'}^{(z)} |_{g_i=const}\right) \right\} =, \tag{22}$$
$$= -\vec{e}_z \left\{ rot\left(j_z \vec{H}_{-s'}^{(z)} |_{g_i=const}\right) + j_z \, rot\left(\vec{H}_{-s'}^{(z)} |_{g_i=const}\right) \right\} = -\vec{e}_z \, rot\left(j_z \vec{H}_{-s'}^{(z)} |_{g_i=const}\right) - \vec{j}_z i\omega\varepsilon_0 \varepsilon \vec{E}_{-s'}^{(z)}.$$

Taking into account that

$$\int\limits_{S_\perp^{(z)}(z)} rot\left(j_z \vec{H}_{-s'}^{(z)}\big|_{g_i=const}\right)\vec{e}_z dS = \int\limits_{S_\perp^{(z)}(z)} rot\left(j_z \vec{H}_{-s'}^{(z)}\big|_{g_i=const}\right)d\vec{S} = \oint j_z \vec{H}_{-s'}^{(z)}\big|_{g_i=const} d\vec{l} = 0, \tag{23}$$

we get

$$N_{s'}^{(z)} \frac{dC_{-s'}}{dz} = -\int\limits_{S_\perp^{(z)}(z)} \left\{ j\vec{E}_{s'}^{(z)} - \sum_s \left\{ C_s\left(\vec{E}_{s'}^{(z)}\vec{H}_s^{(\nabla)} + \vec{H}_{s'}^{(z)}\vec{E}_s^{(\nabla)}\right) + C_{-s}\left(\vec{E}_{s'}^{(z)}\vec{H}_{-s}^{(\nabla)} + \vec{H}_{s'}^{(z)}\vec{E}_{-s}^{(\nabla)}\right) \right\} \right\} dS, \tag{24}$$

$$N_{s'}^{(z)} \frac{dC_{s'}}{dz} = \int\limits_{S_\perp^{(z)}(z)} \left\{ j\vec{E}_{-s'}^{(z)} - \sum_s \left\{ C_s\left(\vec{E}_{-s'}^{(z)}\vec{H}_s^{(\nabla)} + \vec{H}_{-s'}^{(z)}\vec{E}_s^{(\nabla)}\right) + C_{-s}\left(\vec{E}_{-s'}^{(z)}\vec{H}_{-s}^{(\nabla)} + \vec{H}_{-s'}^{(z)}\vec{E}_{-s}^{(\nabla)}\right) \right\} \right\} dS. \tag{25}$$

Separation the exponential dependence on z

$$\vec{E}_k^{(z)}, \vec{H}_k^{(z)} = \tilde{\vec{E}}_k^{(z)}, \tilde{\vec{H}}_k^{(z)} \exp(\gamma_k^{(z)} z) \tag{26}$$

gives

$$\int\limits_{S_\perp^{(z)}(z)} \left(\vec{E}_k^{(z)}\vec{H}_{k'}^{(\nabla)} + \vec{H}_k^{(z)}\vec{E}_{k'}^{(\nabla)}\right) dS = \exp(\gamma_{k'}^{(z)} z + \gamma_k^{(z)} z) W_{k',k}^{(z)} + z \frac{d\gamma_{k'}^{(z)}}{dz} \tilde{N}_{k'}^{(z)}(z) \delta_{k,-k''}, \tag{27}$$

where

$$W_{k',k}^{(z)} = \sum_i \frac{dg_i^{(z)}}{dz} \int\limits_{S_\perp^{(z)}(z)} \left\{ \frac{\partial}{\partial g_i^{(z)}}\left[\tilde{\vec{E}}_{k'}^{(z)} \tilde{\vec{H}}_k^{(z)}\right] - \left[\tilde{\vec{E}}_k^{(z)} \frac{\partial \tilde{\vec{H}}_{k'}^{(z)}}{\partial g_i^{(z)}}\right] - \left[\tilde{\vec{E}}_{k'}^{(z)} \frac{\partial \tilde{\vec{H}}_k^{(z)}}{\partial g_i^{(z)}}\right] \right\} \vec{e}_z dS \tag{28}$$

Let's introduce the new amplitudes

$$C_{\pm s} = \tilde{C}_{\pm s} \exp\left(\mp \gamma_s^{(z)} z \pm \int^z dz' \gamma_s^{(z)}\right), \tag{29}$$

The equations (24) and (25) then transform into

$$N_{s'}^{(z)} \frac{d\tilde{C}_{s'}}{dz} + \sum_s \tilde{C}_s \exp\{-p_{s'}(z) + p_s(z)\} W_{s,-s'}^{(z)} + \sum_s \tilde{C}_{-s} \exp\{-p_{s'}(z) - p_s(z)\} W_{-s',-s}^{(z)} = \exp\{-p_{s'}(z)\} \int\limits_{S_\perp^{(z)}(z)} j\vec{E}_{-s'}^{(z)} dS, \tag{30}$$

$$N_{s'}^{(z)} \frac{d\tilde{C}_{-s'}}{dz} - \sum_s \tilde{C}_{-s} \exp\{p_{s'}(z) - p_s(z)\} W_{-s,s'}^{(z)} - \sum_s \tilde{C}_s \exp\{p_{s'}(z) + p_s(z)\} W_{s',s}^{(z)} = -\exp\{p_{s'}(z)\} \int\limits_{S_\perp^{(z)}(z)} j\vec{E}_{s'}^{(z)} dS, \tag{31}$$

where

$$p_s^{(z)}(z) = \int^z dz' \gamma_s^{(z)}. \tag{32}$$

The system of differential equations (30) and (31) is rigorous and describes the field amplitudes for any non-periodic structured waveguide. For the periodic structured waveguide ($W_{s',s}^{(z)} = 0$) we obtain the well-known system (see, for example, [5,6])

### 3. APPROXIMATE EQUATIONS

In the simplest approach we take into account only one propagating wave ($\gamma_s^{(z)} = ih_s^{(z)}$)

$$N_{s'}^{(z)} \frac{d\tilde{C}_{s'}}{dz} + \tilde{C}_{s'} W_{s',-s'}^{(z)} = \exp\{-p_{s'}(z)\} \int\limits_{S_\perp^{(z)}(z)} j\vec{E}_{-s'}^{(z)} dS. \tag{33}$$

This approach can be used when $\left|\frac{1}{g_i^{(z)}} \frac{dg_i^{(z)}}{dz}\right| \ll 1$.

For propagating wave $\vec{E}_{-s} = \vec{E}_s^*$, $\vec{H}_{-s} = -\vec{H}_s^*$ and we obtain

$$W_{s',-s'}^{(z)} = \frac{1}{2} \frac{dN_{s'}^{(z)}}{dz} + i\Phi_{s'}^{(z)}, \tag{34}$$

where

$$\Phi_{s'}^{(z)} = 2\sum_i \frac{dg_i^{(z)}}{dz} \left\{ -\frac{\partial}{\partial g_i^{(z)}} \operatorname{Im} P_{s'}^{(z)} + \operatorname{Im} \int\limits_{S_\perp^{(z)}(z)} \left[\vec{E}_{s'}^{(z)} \frac{\partial \vec{H}_{s'}^{(z)*}}{\partial g_i^{(z)}}\right] \vec{e}_z dS \right\}, \tag{35}$$

$$N_{s'}^{(z)} = \int\limits_{S_\perp^{(z)}} \left\{ -\left[\vec{E}_{s'}^{(z)*} \vec{H}_{s'}^{(z)}\right] - \left[\vec{E}_{s'}^{(z)} \vec{H}_{s'}^{(z)*}\right] \right\} \vec{e}_z dS = -2\operatorname{Re} \int\limits_{S_\perp^{(z)}} \left[\vec{E}_{s'}^{(z)} \vec{H}_{s'}^{(z)*}\right] \vec{e}_z dS = -4\operatorname{Re} P_{s'}^{(z)}, \tag{36}$$

$$P_{s'}^{(z)} = \frac{1}{2} \int\limits_{S_\perp^{(z)}} \left[\vec{E}_{s'}^{(z)} \vec{H}_{s'}^{(z)*}\right] \vec{e}_z dS. \tag{37}$$

$P_{s'}^{(z)}(z)$ is a complex power that fields $\vec{E}_{s'}^{(z)}$ $\vec{H}_{s'}^{(z)}$ transfer through the cross-section $S_{\perp}^{(z)}(z)$. As for periodic waveguide $\operatorname{Re} P_{s'}$ is a constant value, then the active power $\operatorname{Re} P_{s'}^{(z)}(z)$ is a function of $g_1^{(z)}(z),...,g_I^{(z)}(z)$ only and $\sum_i \frac{dg_i^{(z)}}{dz} \frac{\partial \operatorname{Re} P_{s'}^{(z)}}{\partial g_i^{(z)}} = \frac{d \operatorname{Re} P_{s'}^{(z)}}{dz}$. This is not true about reactive power $\operatorname{Im} P_{s'}^{(z)}$. It can be shown that it is a rapidly changing function with $z$ and $\sum_i \frac{dg_i^{(z)}}{dz} \frac{\partial \operatorname{Im} P_{s'}^{(z)}}{\partial g_i^{(z)}} \neq \frac{d \operatorname{Im} P_{s'}^{(z)}}{dz}$.

In the previous works (see, for example, [7,8,9,10,11,12,13,14]), equations without the second term (phase $\Phi_{s'}^{(z)}$) in (34) were used. In this approach Eq.(33) takes the form

$$\frac{d\tilde{C}_{s'}}{dz} + \frac{1}{2N_{s'}^{(z)}} \frac{dN_{s'}^{(z)}}{dz} \tilde{C}_{s'} = \exp\{-p_{s'}(z)\} \frac{1}{N_{s'}^{(z)}} \int_{S_{\perp}^{(z)}(z)} \vec{j}\vec{E}_{s'}^{(z)*} dS. \tag{38}$$

The eigen vector $\vec{E}_s$ of a periodic waveguide can be represented as

$$\vec{E}_s = \mathcal{E}_0 \sum_n \vec{E}_{s,n}(\vec{r}_\perp) \exp\left(i\frac{2\pi n}{D}z\right). \tag{39}$$

For TH waves with azimuthal symmetry $\tilde{E}_{s,n,z}(0) \neq 0$ and we can assume that $\tilde{E}_{s,0,z}(0) = 1$. Then for a non-periodic waveguide the series impedance that depends on $z$ can be introduced

$$R_{ser}^{(z)}(z) = \frac{\mathcal{E}_0^2}{\operatorname{Re} P_{s'}^{(z)}(z)} \tag{40}$$

The equation (38) can be rewritten in a more usual form ($\tilde{\tilde{C}}_{s'} = \mathcal{E}_0 \tilde{C}_{s'}$) $P_{s'}^{(z)}(z)$

$$\frac{d\tilde{\tilde{C}}_{s'}}{dz} - \frac{1}{2R_{ser}^{(z)}} \frac{dR_{ser}^{(z)}}{dz} \tilde{\tilde{C}}_{s'} = -\exp\{-p_{s'}(z)\} \frac{R_{ser}^{(z)}}{4} \sum_n \exp\left(i\frac{2\pi n}{D(z)}z\right) \int_{S_{\perp}^{(z)}(z)} \vec{j}\vec{E}_{s',n}^{(z)*}(z,r) dS. \tag{41}$$

The vector of electric field in this case is

$$\vec{E}(r,z) = \tilde{C}_s(z)\exp\{p_s(z)\}\vec{E}_s^{(z)}(z,r) + \frac{\vec{j}_z}{i\omega\varepsilon_0\varepsilon} = \tilde{C}_s(z)\exp\{p_s(z)\}\sum_n \vec{E}_{s,n}^{(z)}(z,r)\exp\left(i\frac{2\pi n}{D(z)}z\right) + \frac{\vec{j}_z}{i\omega\varepsilon_0\varepsilon}. \tag{42}$$

A power that electromagnetic fields transfer through the cross-section $S_{\perp}^{(z)}(z)$ of the lossless waveguide without the electron beam ($\vec{j} = 0$) proportional to the product of two factors $\left|\tilde{C}_s(z)\right|^2 \operatorname{Re} P_{s'}^{(z)}(z) \sim R_{ser}^{(z)}(z) \operatorname{Re} P_{s'}^{(z)}(z)$ and is a constant value (see (40)

A single-wave approach without an additional "phase" term $i\Phi_{s'}^{(z)}$ is widely used to calculate the characteristics of non-periodic accelerating and other slow wave structures. A procedure of calculation of $\Phi_{s'}^{(z)}$ is not simple, especially its second part. The role of this term and the conditions when it can be neglected will be studied in the future work.

It should be noted that in the approach proposed above the change in the phase of the electric field inside one cell $p_s^{(z)}(z) = i\int_0^z dz' h_s^{(z)}$ has a complicated character as the local wave vector $h_s^{(z)}(z) = h_s^{(z)}\left(g_1^{(z)}(z),...,g_I^{(z)}(z)\right)$ depends on several functions.

Moreover, the additional dependence of space harmonic exponents on $z$ ($\exp(i2\pi n/D(z))$) can arise if the longitudinal characteristics of waveguide change along the $z$ axis.

## 2. CONCLUSIONS

In this work a new generalization of the theory of coupled modes for non-periodic structured waveguide is presented. Based on a set of eigen waves of a homogeneous periodic waveguide, a new basis of vector functions is introduced that takes into account the non-periodicity of the waveguide. Representing the total field as the sum of these functions with unknown scalar coefficients, a system of coupled equations that determines the dependence of these coefficients on the longitudinal coordinate have been obtained. A single-wave equation in proposed model has an additional "phase" term that was not used previously. In the frame of proposed approach, the z-dependent series impedance and local wave vector were introduced for structured waveguides

The next improvement of the existing single-wave models can be made by taking into account the backward wave. It can be useful for understanding the procedure for tuning inhomogeneous accelerating structures. In all tuning methods used, the focus is on obtaining a given phase shift between cells with an accuracy of less than one degree. But for the non-periodic waveguide there is always the backward wave. Therefore, together with forward wave in the waveguide there is a standing wave with a small amplitude. Can we get a uniform distribution of phases in this case? And if it possible, what kind of amplitude distribution do we create?

Couple-mode approach needs to know the dependence of characteristics of eigen modes of homogeneous waveguide on some geometrical parameters. Since the calculation of the modes of periodic waveguides is not a simple task, the matrix approach, which provides a procedure for calculation of the distribution of the electric field, seems to be useful for simple geometries [15,16].

## ACKNOWLEDGEMENTS

The author would like to thank David Reis and Valery Dolgashev for their support.